%% file: ChiralKondoZL_Main.tex
\documentclass[english,aps,prl, twocolumn,longbibliography]{revtex4-2}
\usepackage[T1]{fontenc}
\usepackage[latin9]{inputenc}
\usepackage{color}
\usepackage{babel}
\usepackage{float}
\usepackage{wrapfig}
\usepackage{bm}
\usepackage{amsmath}
\usepackage{amssymb}
\usepackage{graphicx}
\usepackage[unicode=true,pdfusetitle,
 bookmarks=true,bookmarksnumbered=false,bookmarksopen=false,
 breaklinks=false,pdfborder={0 0 0},pdfborderstyle={},backref=false,colorlinks=true]
 {hyperref}

\makeatletter

\providecommand{\tabularnewline}{\\}

\hypersetup{colorlinks,allcolors=blue}

\usepackage[caption=false]{subfig}
\newcommand{\phantomsubfloat}[1]{
    {
        \captionsetup[subfigure]{labelformat=empty}
        \subfloat[][]{#1}     }%
} 
\usepackage{braket}
\usepackage{bbold}
\usepackage{xr}
\makeatother

\begin{document}
\title{Manipulating Non-Abelian Anyons in a Chiral Multichannel Kondo Model}
\author{Matan Lotem}
\email{matanlotem@mail.tau.ac.il}

\author{Eran Sela}
\email{eransx@googlemail.com}

\author{Moshe Goldstein}
\email{mgoldstein@tauex.tau.ac.il}

\affiliation{Raymond and Beverly Sackler School of Physics and Astronomy, Tel Aviv
University, Tel Aviv 6997801, Israel}
\begin{abstract}
Non-Abelian anyons are fractional excitations of gapped topological
models believed to describe certain topological superconductors or
quantum Hall states. Here, we provide the first numerical evidence
that they emerge as independent entities also in gapless electronic
models. Starting from a multi-impurity multichannel chiral Kondo model,
we introduce a novel mapping to a single-impurity model, amenable
to Wilson's numerical renormalization group. We extract its spectral
degeneracy structure and fractional entropy, and calculate the $F$
matrices, which encode the topological information regarding braiding
of anyons, directly from impurity spin-spin correlations. Impressive
recent advances on realizing multichannel Kondo systems with chiral
edges may thus bring anyons into reality sooner than expected.
\end{abstract}
\maketitle
\global\long\def\k#1{\Ket{#1}}%
\global\long\def\b#1{\Bra{#1}}%
\global\long\def\bk#1{\Braket{#1}}%
\renewcommand{\figurename}{FIG.}
\renewcommand{\tablename}{TABLE.}

\paragraph{Introduction.---}

Non-Abelian anyons are exotic (quasi-)particles which obey neither
fermionic nor bosonic statistics, and lie at the heart of topological
quantum computing \citep{kitaevFaulttolerantQuantumComputation2003,nayakNonAbelianAnyonsTopological2008}.
They define an anyonic fusion space which can only be transversed
by their mutual exchange, or braiding, thus providing topological
protection for information encoded in this space. An important class
of non-Abelian anyons are the $\mathrm{SU}\!\left(2\right)_{k}$ anyons,
which are governed by truncated $\mathrm{SU}\!\left(2\right)$ fusion
rules \citep{bondersonNonAbelianAnyonsInterferometry2007}. Each such
anyon (of topological charge $\frac{1}{2}$) carries with it a quantum
dimension of $d_{k}{=}2\cos(\frac{\pi}{2+k})$, which gives the degeneracy
per anyon in the thermodynamic limit. Prominent examples are the Ising
($k{=}2,d_{2}{=}\sqrt{2}$) and Fibonacci ($k{=}3,d_{3}{=}\frac{1+\sqrt{5}}{2}$)
anyons, predicted to arise, e.g., in the $\nu{=}\frac{5}{2}$ and
$\nu{=}\frac{12}{5}$ fractional quantum Hall states, respectively
\citep{mooreNonabelionsFractionalQuantum1991,readPairedQuantumHall1999},
and Majorana ``fermions'' (also $k{=}2$), which arise in a variety
of topological systems, e.g., pinned to vortices in 2D topological
superconductors \citep{readPairedStatesFermions2000,ivanovNonAbelianStatisticsHalfQuantum2001,fuSuperconductingProximityEffect2008}
or on the edges of superconducting nanowires \citep{lutchynMajoranaFermionsTopological2010,oregHelicalLiquidsMajorana2010}.
However, these quasiparticles prove to be extremely elusive, with
no clear experimental evidence for their non-Abelian nature.

Another system governed by $\mathrm{SU}\!\left(2\right)_{k}$ fusion
rules, although not of a topological nature, is the $k$-channel Kondo
effect \citep{nozieresKondoEffectReal1980,hewsonKondoProblemHeavy1993}.
This was most clearly demonstrated by \citet{emeryMappingTwochannelKondo1992},
who formulated the solution of the two-channel Kondo effect in terms
of Majorana operators. Importantly, this effect has already been observed
in tunable nanostructures, for both $k{=}2$ \citep{potokObservationTwochannelKondo2007,mebrahtuQuantumPhaseTransition2012,mebrahtuObservationMajoranaQuantum2013,kellerUniversalFermiLiquid2015,iftikharTwochannelKondoEffect2015}
and $k{=}3$ \citep{iftikharTunableQuantumCriticality2018} channels.
The Kondo effect occurs when a quantum impurity, e.g., a spin-$\frac{1}{2}$,
is coupled antiferromagnetically to (multiple) noninteracting spinfull
fermionic bath(s), i.e., channel(s). For a single channel, at temperatures
below the Kondo temperature, the fermions in the bath screen the impurity,
which can be interpreted as the impurity binding a fermion from the
bath and forming a singlet with it. Going to multiple channels, each
channel independently contributes a single screening fermion, but
this leads to frustration and fractionalization of the impurity degrees
of freedom. The fractionalized quasiparticle comes with a zero-temperature
entropy of $\log d_{k}$, corresponding to the quantum dimension of
a single $\mathrm{SU}\!\left(2\right)_{k}$ charge-$\frac{1}{2}$
anyon \citep{tsvelickThermodynamicsMultichannelKondo1985}. Indeed,
the low-energy physics of the $k$-channel spin-$s\leq\frac{k}{2}$
Kondo effect are captured by a conformal field theory (CFT) in which
a single $\mathrm{SU}\!\left(2\right)_{k}$ anyon with charge $s$
is fused onto the primary fields of ($k$-channel) free fermions \citep{affleckKondoEffectConformal1991,affleckCriticalTheoryOverscreened1991}.

In order to discuss anyonic statistics, or braiding, we require (i)
multiple quasiparticles, and (ii) a physically accessible operator
which acts on the anyonic fusion space. The paradigmatic multichannel
Kondo effect assumes a dilute scenario, so that at temperatures above
the Fermi velocity over the interimpurity separation ($v_{F}/R$),
each impurity is effectively coupled to a different bath, thus satisfying
(i) but breaking (ii), while for lower temperatures, the bath fermions
mediate effective RKKY \citep{rudermanIndirectExchangeCoupling1954,kasuyaTheoryMetallicFerro1956,yosidaMagneticPropertiesCuMn1957}
interactions between the impurities, thus resolving the frustration
and avoiding emergent fractionalized quasiparticles. It was only recently
realized that (i) and (ii) might be reconciled, either by gapping
out the bath via superconducting pairing \citep{komijaniIsolatingKondoAnyons2020}
or preventing the generation of interactions in the first place by
employing chiral channels \citep{lopesAnyonsMultichannelKondo2020}.
In the latter, fermions (of all channel and spin species) can propagate
only in one direction, as on the edge of an integer quantum Hall system,
thus preventing backscattering and interference, the mechanisms behind
effective interactions. Intuitively, the first impurity encountered
by chiral fermions is unaware of the impurities to follow, thus fractionalizing
as in the single-impurity case. Repeating this argument sequentially
suggests a fractionalized quasiparticle for each impurity. \citet{lopesAnyonsMultichannelKondo2020}
introduced a multiple-impurity extension of the single-impurity multichannel
Kondo CFT fusion as an ansatz for the low-energy behavior of such
a system: for each spin-$\frac{1}{2}$ impurity introduce an $\mathrm{SU}\!\left(2\right)_{k}$
anyon with ``topological'' charge $\frac{1}{2}$, fuse these anyons
to each other, defining a non-Abelian fusion space, and then fuse
the result onto the free-fermionic primary fields (see examples in
Sec.\,\ref{sec:CFT} of the Supplemental Material \citep{SM}).
In this ansatz, different fusion outcomes (corresponding to different
states in the fusion space) leave signatures, e.g., on the spatial
fermionic correlation functions, which (in principle) can be measured
by interferometry, enabling measurement-only braiding \citep{bondersonMeasurementOnlyTopologicalQuantum2008}
of quasiparticles. However, in the CFT ansatz the anyons were put
in by hand.

In this Letter we independently test this conjecture, employing a
controlled, nonperturbative, numerically exact method---Wilson's
numerical renormalization group (NRG) \citep{wilsonRenormalizationGroupCritical1975},
which enables zooming in on the low-energy physics of quantum impurity
problems. A key part of NRG is mapping the bath onto a tight-binding
(Wilson) chain, but this is incompatible with chirality, as any notion
of direction in a (nearest-neighbor) tight-binding chain can be absorbed
by gauge transformations. However, chirality is also the solution
to the problem. As the distance between the impurities typically enters
through interference effects, which are now forbidden, we argue that
it does not affect universal properties. This is supported by the
results in Ref.\,\citep{lotemChiralNumericalRenormalization2022},
in which we numerically account for the distance, as well as by the
Bethe-ansatz solution for the Kondo problem \citep{andreiSolutionKondoProblem1983}.
We thus have the freedom to take the distance between the impurities
to be arbitrarily small, as long as we retain the notion of chirality
and the ordering of the impurities. We do this by first introducing
``buffer sites'' between the impurities and the bulk chiral channels,
and only then taking the interimpurity distance to zero. This results
in a large effective impurity coupled to a trivial bath, which can
readily be plugged into NRG. We then numerically demonstrate that
the low-energy behavior of the system indeed corresponds to an $\mathrm{SU}\!\left(2\right)_{k}$
charge-$\frac{1}{2}$ anyon for each impurity, and that the fusion
outcome of pairs of such anyons can be probed by measuring interimpurity
spin correlations.

\paragraph{Model and method.---}

\begin{figure}
\begin{centering}
\includegraphics[width=1\columnwidth]{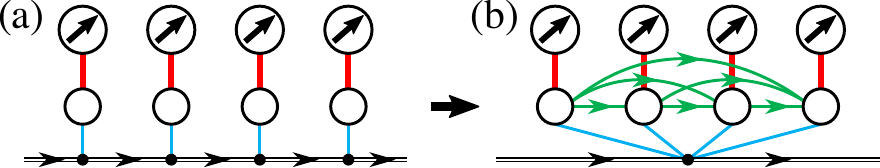}
\par\end{centering}
\begin{centering}
\phantomsubfloat{\label{fig:Dangling-Model}}\phantomsubfloat{\label{fig:Effective-Model}}
\par\end{centering}
\vspace{-2.5em}
\caption{(a) The impurities are Kondo-coupled to \textquotedblleft buffer\textquotedblright{}
dangling sites, which in turn quadratically couple to the chiral channels,
and are considered part of the noninteracting bath. (b) Taking the
distance between these sites to zero leads to an effective chiral
model, in which the dangling sites together with the impurities form
a large effective impurity.}
\end{figure}

We start with $M$ spin-$\frac{1}{2}$ impurities with spin operator
$\mathbf{S}_{m}$ where $m\in\left\{ 1,...,M\right\} $, and a bath
of right-moving free fermions 
\begin{align}
H_{\mathrm{chiral}} & =\sum_{\alpha\sigma}\int dx\psi_{\alpha\sigma}^{\dagger}\left(x\right)\left(-iv_{F}\partial_{x}\right)\psi_{\alpha\sigma}\left(x\right),\label{eq:H-chiral}
\end{align}
with Fermi velocity $v_{F}$, spin $\sigma\in\left\{ \uparrow,\downarrow\right\} $,
and channel $\alpha\in\left\{ 1,..,k\right\} $. One can directly
couple the impurities to the bath at locations $\left\{ R_{m}\right\} $
by writing the Hamiltonian $\sum_{m}J\mathbf{S}_{m}\cdot\mathbf{s}\left(R_{m}\right)+H_{\mathrm{chiral}}$,
with $J>0$ the Kondo coupling and $\mathbf{s}\left(x\right)\equiv\sum_{\alpha\sigma}\psi_{\alpha\sigma}^{\dagger}\left(x\right)\bm{{\sigma}}_{\sigma\sigma^{\prime}}\psi_{\alpha\sigma^{\prime}}\left(x\right)$
the bath spin at location $x$. We treat such a model in Ref.\,\citep{lotemChiralNumericalRenormalization2022}
by introducing $M$ coupled effective $k$-channel baths, but this
comes with a very high computational price tag, due to the exponential
scaling of NRG with the number of channels. Instead, here we employ
a mapping which captures the chirality with a single $k$-channel
bath. We first separate the impurities from the bath, as illustrated
in Fig.\,\ref{fig:Dangling-Model}, by introducing buffer ``dangling''
fermionic sites coupled to the bath at locations $\left\{ R_{m}\right\} $,
and then couple the impurities to these dangling sites, arriving at
\begin{align}
H_{\mathrm{total}} & =J\sum_{m}\mathbf{S}_{m}\cdot\mathbf{s}_{m}+H_{\mathrm{dang}}+H_{\mathrm{chiral}},\label{eq:H-total}\\
H_{\mathrm{dang}} & =\tilde{t}_{0}\!\sum_{m\alpha\sigma}\!\left[d_{m\alpha\sigma}^{\dagger}\psi_{\alpha\sigma}\!\left(R_{m}\right)\!+\psi_{\alpha\sigma}^{\dagger}\!\left(R_{m}\right)\!d_{m\alpha\sigma}\right],\label{eq:H-dangling}
\end{align}
where $d_{m\alpha\sigma}$ and $\mathbf{s}_{m}\equiv\sum_{\alpha\sigma}d_{m\alpha\sigma}^{\dagger}\bm{{\sigma}}_{\sigma\sigma^{\prime}}d_{m\alpha\sigma^{\prime}}$
are the dangling-site fermionic and spin operators, respectively,
$J>0$ is the Kondo coupling, and $\tilde{t}_{0}$ together with the
Fermi velocity define a soft cutoff $\Gamma\equiv\frac{\tilde{t}_{0}^{2}}{2v_{F}}$.

Initially we treat the dangling sites together with the chiral channels
as the noninteracting bath to which the impurities are coupled. As
typical of Kondo problems, the bath dependence of impurity quantities
enters (to all orders in the Kondo coupling $J$) only through the
(retarded) Green function of the bath at the sites coupled to the
impurities, i.e., the dangling sites, when these are decoupled from
the impurities: 
\begin{equation}
\mathbf{g}_{\mathrm{dang}}^{R}\left(\omega\right)=\left[\omega\mathbb{1}-\mathbf{h}-\mathbf{\Sigma}^{R}\left(\omega\right)\right]^{-1},
\end{equation}
with $\mathbb{1}$ the $M{\times}M$ identity matrix, $\mathbf{h}{=}0$
the single-particle Hamiltonian acting on the dangling sites, and
\begin{equation}
\mathbf{\Sigma}_{mm^{\prime}}^{R}\left(\omega\right)=-2i\Gamma\Theta\left(R_{m^{\prime}}-R_{m}\right)e^{i\omega\left(R_{m^{\prime}}-R_{m}\right)/v_{F}},
\end{equation}
the retarded self-energy due to the coupling of the dangling sites
to the chiral channels, where $\Theta\left(x\right)$ is the Heaviside
step function {[}taking $\Theta\left(0\right)=\frac{1}{2}${]}. A
clear signature of chirality (assuming right movers) is that any retarded
quantity at location $r$ due to an event at $r^{\prime}>r$ vanishes.
And indeed, all elements below the diagonal of $\mathbf{\Sigma}^{R}\left(\omega\right)$
are zero, as a result of which the same holds for $\mathbf{g}_{\mathrm{dang}}^{R}\left(\omega\right)$.
Thus, importantly, the introduction of the dangling sites retains
chirality. The obtained model is formally equivalent to one without
dangling sites in the $\Gamma\to\infty$ limit, whereas for finite
$\Gamma$ we have merely modified the bath density of states to a
Lorentzian of width $\Gamma$ at each dangling site, which should
not affect the universal low-energy properties. Assuming $J<\Gamma$,
we can define the Kondo temperature as $T_{K}=\Gamma e^{-\pi\Gamma/J}$.

We now take the limit $\omega\!\left(R_{M}{-}R_{1}\right)\!/v_{F}\to0$,
corresponding to low temperatures or long wavelengths. This limit
is taken after the infinite bandwidth limit of Eq.\,(\ref{eq:H-chiral}),
and is not impaired by the soft cutoff $\Gamma$. $\mathbf{\Sigma}^{R}\left(\omega\right)$
loses its frequency dependence, but not its chirality, and can be
written as
\begin{equation}
\mathbf{\Sigma}_{mm^{\prime}}^{R}\to-i\Gamma\!\left\{ \!\!\begin{smallmatrix}2 & m^{\prime}>m\\
1 & m^{\prime}=m\\
0 & m^{\prime}<m
\end{smallmatrix}\right.\equiv\mathbf{h}_{mm^{\prime}}^{\mathrm{eff}}-i\Gamma,\label{eq:g-effecitve}
\end{equation}
with $\mathbf{h}^{\mathrm{eff}}$ a Hermitian matrix. Thus, $\mathbf{h}^{\mathrm{eff}}$
can be interpreted as an effective (single-particle) Hamiltonian coupling
all dangling sites to each other via imaginary hopping amplitudes,
while $-i\Gamma$ describes a single trivial bath coupled equally
to all dangling sites, i.e.,
\begin{align}
H_{\mathrm{dang}}^{\mathrm{eff}} & =\sum_{\alpha\sigma}\sum_{m>m^{\prime}}it_{mm^{\prime}}^{\prime}\left[d_{m\alpha\sigma}^{\dagger}d_{m^{\prime}\alpha\sigma}-d_{m^{\prime}\alpha\sigma}^{\dagger}d_{m\alpha\sigma}\right]\nonumber \\
 & +\sqrt{M}\tilde{t}_{0}\sum_{m\alpha\sigma}\left[d_{m\alpha\sigma}^{\dagger}\psi_{\alpha\sigma}\!\left(0\right)\!+\psi_{\alpha\sigma}^{\dagger}\!\left(0\right)\!d_{m\alpha\sigma}\right],\label{eq:H-effective}
\end{align}
with $t_{mm^{\prime}}^{\prime}=\Gamma$. Replacing $H_{\mathrm{dang}}$
in Eq.\,(\ref{eq:H-total}) with $H_{\mathrm{dang}}^{\mathrm{eff}}$,
we arrive at the model depicted in Fig.\,\ref{fig:Effective-Model}.

Let us review what we have achieved. The obtained model is still chiral
(for the very specific choice of $t_{mm^{\prime}}^{\prime})$, and
reproduces the bath Green function in the low temperature limit. But
now we can interpret the impurities together with the dangling sites
as a large effective impurity, coupled to an effective bath (described
only by $H_{\mathrm{chiral}}$) at a single location, so that its
chirality is no longer important. The resulting structure also hints
at first fusing all the impurities together, and then fusing onto
a single (multichannel) bath, as in the CFT ansatz of Ref.\,\citep{lopesAnyonsMultichannelKondo2020}.
The obtained model is amendable to standard NRG, although one still
needs to account for the multiple channels. In order to reduce the
computational cost, we exploit the different symmetries of the model
(charge, spin, channel), using the \textsc{QSpace} tensor network
library, which treats Abelian and non-Abelian symmetries on equal
footing \citep{weichselbaumTensorNetworksNumerical2012,weichselbaumNonabelianSymmetriesTensor2012,weichselbaumXsymbolsNonAbelianSymmetries2020}.
For implementation details see Sec.\,\ref{sec:NRG-Details} in the
Supplemental Material \citep{SM} and Refs.\,\citep{bullaNumericalRenormalizationGroup2008,campoAlternativeDiscretizationNumerical2005,zitkoAdaptiveLogarithmicDiscretization2009,sakuraiModernQuantumMechanics2017b}
therein. In order to apply NRG, we introduce an artificial sharp high-energy
cutoff \textbf{$D\gg\Gamma,J$ }to the bath density of states. This
cutoff, and to a lesser extent the NRG discretization and truncation,
mimic the effect of the bulk bands (Landau levels), setting a finite
bandwidth for the chiral edge mode, and mediating effective nonchiral
RKKY interactions between the impurities. The latter are expected
to decay exponentially with both the bulk gap and the interimpurity
distance \citep{bloembergenNuclearSpinExchange1955,kurilovichIndirectExchangeInteraction2016,kurilovichIndirectExchangeInteraction2017},
and are thus eliminated by numerically tuning each $t_{mm^{\prime}}^{\prime}$
slightly away from $\Gamma$ to reinstate chirality (see Sec.\,\ref{subsec:Fine-Tuning-to-Chirality}
in the Supplemental Material \citep{SM}).

\paragraph{Results.---}

\begin{figure}
\begin{centering}
\includegraphics[width=1\columnwidth]{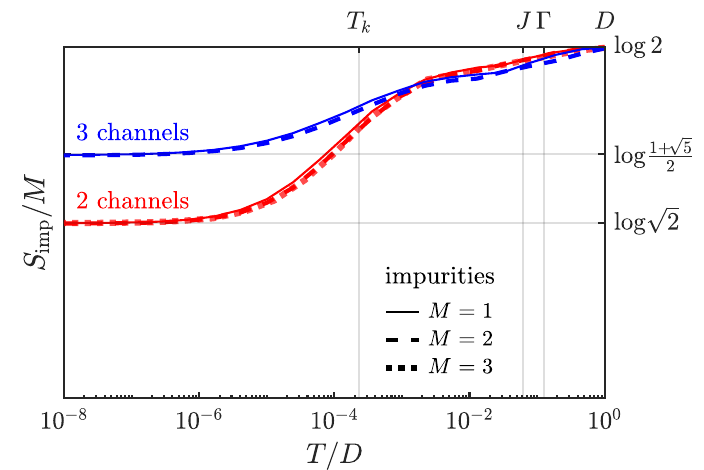}
\par\end{centering}
\caption{Impurity entropy per impurity for two channels (red) with 1--3 impurities,
and three channels (blue) with 1--2 impurities, taking $2J{=}\Gamma{=}D/8$.
At high temperatures the impurity spins are free, each contributing
an entropy of $\log2$. At low temperatures each impurity contributes
a fractional entropy corresponding to the quantum dimension of Ising
{[}$\mathrm{SU}\!\left(2\right)_{2}${]} or Fibonacci {[}$\mathrm{SU}\!\left(2\right)_{3}${]}
anyons for two or three channels, respectively. \label{fig:Impurity-entropy}}
\end{figure}

We apply NRG to the effective Hamiltonian for two channels with up
to three impurities, and for three channels with up to two impurities.
In Fig.\,\ref{fig:Impurity-entropy} we plot the impurity entropy
$S_{\mathrm{imp}}$, defined as the difference between the entropy
of the full system and that of the fermionic bath (dangling sites
+ chiral channels) in the absence of the impurities, which quantifies
the effective degree of freedom $d_{\mathrm{eff}}$ each impurity
introduces. We find that $d_{\mathrm{eff}}$ is independent of the
number of impurities $M$, so that $S_{\mathrm{imp}}/M=\log d_{\mathrm{eff}}\left(k,T\right)$
follows the universal single-impurity curve, matching the limit of
infinitely separated impurities, and thus supporting our argument
that in a chiral system the interimpurity distance is not important.
At high temperatures each impurity is effectively a free spin, contributing
a $d_{\mathrm{eff}}{=}2$ degree of freedom. Going below the Kondo
temperature while assuming the thermodynamic limit for the bath, each
impurity contributes a fractional degree of freedom $d_{\mathrm{eff}}{=}d_{k}$
corresponding exactly to an $\mathrm{SU}\!\left(2\right)_{k}$ anyon.
These results are well known in the single-impurity scenario \citep{tsvelickThermodynamicsMultichannelKondo1985},
but the scaling to multiple impurities, implying an anyon for each
impurity, is quite remarkable. This is very different from the paradigmatic
multi-impurity multichannel scenario, where the initially similar
entropy curves break for temperatures below $\sim v_{F}/R$ due to
coherent backscattering which generates effective RKKY interactions.
In order to probe anyonic statistics we need coherence, and indeed
in our case we are already in the regime of $T\ll v_{F}/R\to\infty$,
but now due to chirality, backscattering is forbidden, and the anyons
survive.

The curves in Fig.\,\ref{fig:Impurity-entropy} were obtained for
the specific choice of the dangling-site hopping amplitudes $t_{mm^{\prime}}^{\prime}$
which renders the system chiral. We can characterize this point by
artificially tuning away from it, and demonstrate that at the chiral
point, the low-energy theory is exactly that of the CFT ansatz of
Ref.\,\citep{lopesAnyonsMultichannelKondo2020}. This is best observed
in the finite-size spectrum obtained by NRG, but as its analysis is
quite technical, we defer it to Sec.\,\ref{sec:Finite-Size-Spectrum}
in the Supplemental Material \citep{SM}. Instead, here we discuss
more intuitive quantities.

\begin{figure}
\begin{centering}
\includegraphics[width=1\columnwidth]{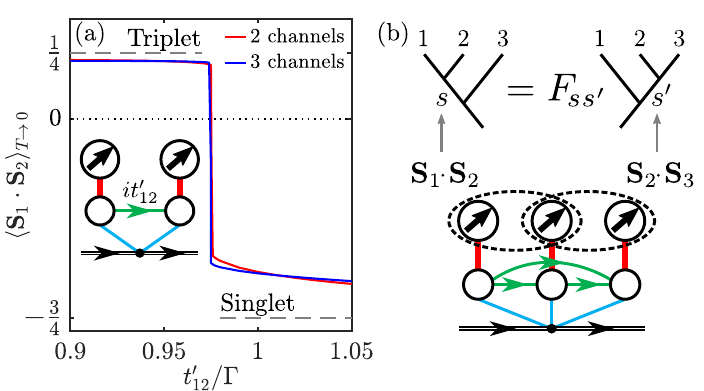}
\par\end{centering}
\begin{centering}
\phantomsubfloat{\label{fig:QPT}}\phantomsubfloat{\label{fig:F-matrix}}\vspace{-2.5em}
\par\end{centering}
\caption{(a) Quantum phase transition for two impurities with two (red) and
three (blue) channels as a function of the dangling-site hopping amplitude
$t_{12}^{\prime}$, taking $2J{=}\Gamma{=}D/8$. Correlations for
the bare singlet and triplet are indicated by dashed lines. (b) Extraction
of the $F$ matrix from interimpurity spin correlators in a three-impurity
system.}
\end{figure}

For two impurities, with either two or three channels, we find that
the effective system undergoes a quantum phase transition from a Kondo-screened
spin-1 impurity when the single parameter $t_{12}^{\prime}$ is below
some critical value to a spin-0 ``Kondo'' effect above it, similar
to the two-impurity Kondo-RKKY phase transition \citep{jayaprakashTwoImpurityKondoProblem1981}.
The two phases can be identified by their low-energy spectra (see
Sec.\,\ref{sec:Finite-Size-Spectrum} in the Supplemental Material
\citep{SM}), with the transition observed, e.g., in the interimpurity
spin correlator $\left\langle \mathbf{S}_{1}\!\cdot\mathbf{S}_{2}\right\rangle _{T\to0}$,
which flips sign from positive (tripletlike) to negative (singletlike),
as shown in Fig.\,\ref{fig:QPT}. Tuning away from criticality and
projecting the operator $\mathbf{S}_{1}\!\cdot\mathbf{S}_{2}$ down
to the low-energy subspace, we find it is a constant (equal to $\left\langle \mathbf{S}_{1}\!\cdot\mathbf{S}_{2}\right\rangle _{T\to0}$),
and thus commutes with the low-energy Hamiltonian. This is consistent
with our characterization of the two phases, but is not trivial, as
$\mathbf{S}_{1}\!\cdot\mathbf{S}_{2}$ does not commute with the full
Hamiltonian, and hence the definite spin states (singlet and triplet)
mix low- and high-energy states. The critical $t_{12}^{\prime}$ is
exactly the hopping amplitude required for the system to be chiral
(it indeed converges to $\Gamma$ for $D\gg\Gamma,J$; see Fig.\,\ref{fig:chiral-hopping}
in the Supplemental Material \citep{SM}). The projected $\mathbf{S}_{1}\!\cdot\mathbf{S}_{2}$
also commutes with the low-energy Hamiltonian at this point, but now
has two eigenvalues, positive and negative. Projecting onto the subspace
corresponding to the negative (positive) eigenvalue takes us back
to the spin-0 (spin-1) Kondo phase. Thus, at the chiral point, the
low-energy Hamiltonian is the direct sum of the low-energy Hamiltonians
of the spin-0 and spin-1 Kondo effects. Remembering these can be obtained
by fusing an $\mathrm{SU}\!\left(2\right)_{k}$ charge-0 or 1 anyon
to the $k$-channel bath, we see that in the chiral case we fuse two
charge-$\frac{1}{2}$ anyons to the bath
\[
0\times\mathrm{Bath}+1\times\mathrm{Bath}=\left(0+1\right)\times\mathrm{Bath}=\tfrac{1}{2}\times\tfrac{1}{2}\times\mathrm{Bath},
\]
in perfect agreement with the CFT ansatz of Ref.\,\citep{lopesAnyonsMultichannelKondo2020}.
As a byproduct we have also demonstrated that a (low-energy) measurement
of the spin correlator $\mathbf{S}_{1}\!\cdot\mathbf{S}_{2}$ actually
measures the fusion outcome of the two anyons. We note that this relation
between the fusion channel and the spin correlator was also recently
demonstrated analytically in the limits of $k{=}2$ and large-$k$
channels \citep{gabayMultiimpurityChiralKondo2022}.

This suggests we can extract the anyonic $F$ matrix, which fully
characterizes the non-Abelian part of the anyonic theory \citep{bondersonNonAbelianAnyonsInterferometry2007},
from measurements of different pairwise spin correlators, as depicted
in Fig.\,\ref{fig:F-matrix}. We explicitly demonstrate this for
three impurities and two channels. We now tune two parameters: the
nearest-neighbor $t_{12}^{\prime}=t_{23}^{\prime}$ (equal by symmetry)
and next-nearest-neighbor $t_{13}^{\prime}$ hopping amplitudes. For
general values the effective low-energy Hamiltonian is that of a single
spin-$\frac{1}{2}$ two-channel Kondo (2CK) effect, $H_{\mathrm{2CK}}$.
However, at a single critical point, corresponding to the system being
chiral, we get a twofold degeneracy (for each energy eigenstate) on
top of this 2CK effect. We can thus write the low-energy Hamiltonian
as a direct sum of two 2CK low-energy Hamiltonians $H_{\mathrm{2CK}}{\oplus}H_{\mathrm{2CK}}$,
each given by CFT by fusing a charge-$\frac{1}{2}$ anyon to the bath
\[
\left(\tfrac{1}{2}+\tfrac{1}{2}\right)\times\mathrm{Bath}=\tfrac{1}{2}\times\left(0+1\right)\times\mathrm{Bath}=\tfrac{1}{2}\times\tfrac{1}{2}\times\tfrac{1}{2}\times\mathrm{Bath}.
\]
This is equivalent to fusing three charge-$\frac{1}{2}$ anyons to
the bath, again in perfect agreement with the CFT ansatz of Ref.\,\citep{lopesAnyonsMultichannelKondo2020}.
We see that the degeneracy is associated to a decoupled fusion space,
and can write the low-energy Hamiltonian as an outer product $H_{\mathrm{2CK}}{\otimes}\mathbb{1}_{2\times2}$,
acting on the ``energy space'' and (trivially) on the fusion space.

Projecting the three pairwise spin correlators $\mathbf{S}_{1}\negmedspace\cdot\mathbf{S}_{2}$,
$\mathbf{S}_{2}\negmedspace\cdot\mathbf{S}_{3}$, and $\mathbf{S}_{1}\negmedspace\cdot\mathbf{S}_{3}$
down to the low-energy subspace, we find all three commute with the
low-energy Hamiltonian, and act nontrivially only on the fusion space.
Thus, for each pair of impurities $m,m^{\prime}$ the projected $\mathbf{S}_{m}\negmedspace\cdot\mathbf{S}_{m^{\prime}}$
can be written as $\mathbb{1}_{\mathrm{2CK}}\otimes\mathbf{s}_{mm^{\prime}}$,
where $\mathbb{1}_{\mathrm{2CK}}$ is the identity matrix in the ``energy
space'' and $\mathbf{s}_{mm^{\prime}}$ is a $2{\times}2$ Hermitian
matrix. Diagonalizing $\mathbf{s}_{mm^{\prime}}$ we find that it
(and thus \textbf{$\mathbf{S}_{m}\negmedspace\cdot\mathbf{S}_{m^{\prime}}$})
has one negative (singletlike) and one positive (tripletlike) eigenvalue,
with eigenstates $\k{0_{mm^{\prime}}}$ and $\k{1_{mm^{\prime}}}$,
respectively. The different correlators do not commute with each other,
and so define different bases for the fusion space, related by the
basis transformation
\begin{equation}
\!F\,{=}\begin{pmatrix}\bk{0_{12}|0_{23}} & \bk{0_{12}|1_{23}}\\
\bk{1_{12}|0_{23}} & \bk{1_{12}|1_{23}}
\end{pmatrix}{=}\frac{1}{\sqrt{2}}\!\begin{pmatrix}1.003 & 0.997\\
0.997 & -1.003
\end{pmatrix}\!.\label{eq:F-matrix}
\end{equation}
For concreteness we have restricted ourselves to the relation between
the eigenbases of $\mathbf{S}_{1}\negmedspace\cdot\mathbf{S}_{2}$
and $\mathbf{S}_{2}\negmedspace\cdot\mathbf{S}_{3}$, and presented
the numerically extracted values in this case. We note that this result
displays dependence on the ratio $J/\Gamma$, which we discuss in
Sec.\,\ref{sec:Parameters} of the Supplemental Material \citep{SM}.
Interpreting the eigenstates of the spin correlator $\mathbf{S}_{m}\negmedspace\cdot\mathbf{S}_{m^{\prime}}$
as states with definite fusion outcomes of anyons $m$ and $m^{\prime}$
(as in the two-impurity case), Eq.\,(\ref{eq:F-matrix}) exactly
defines the $F$ matrix, which matches $\frac{1}{\sqrt{2}}\!\!\left(\begin{smallmatrix}1 & 1\\
1 & -1
\end{smallmatrix}\!\right)$ corresponding to $\mathrm{SU}\!\left(2\right)_{2}$ anyons.

\paragraph{Conclusions.---}

We have numerically demonstrated that multiple Kondo impurities coupled
to $k$ chiral channels (i) host multiple $\mathrm{SU}\!\left(2\right)_{k}$
non-Abelian anyons (one per impurity), highlighted by the fractional
entropy contribution per impurity, and (ii) the emergence of a decoupled
fusion space, which can be probed by low-energy measurements of the
interimpurity spin correlators, explicitly extracting the $F$ matrix
of $\mathrm{SU}\!\left(2\right)_{2}$ anyons. The anyons can now be
braided by a measurement-only protocol \citep{bondersonMeasurementOnlyTopologicalQuantum2008},
which teleports them using only measurements of pairwise topological
charge (fusion channel). One can envision implementing this protocol,
e.g., by a low-energy scattering experiment, directly demonstrating
the non-Abelian nature of the anyons in the system.

Experiments consisting of a single impurity coupled to two and three
integer quantum Hall edge states (i.e., chiral channels) have already
been carried out \citep{iftikharTwochannelKondoEffect2015,iftikharTunableQuantumCriticality2018},
with clear signatures of the fractionalized degrees of freedom \citep{landauChargeFractionalizationTwoChannel2018,vandalumWiedemannFranzLawNonFermi2020,nguyenThermoelectricTransportThreeChannel2020,hanFractionalEntropyMultichannel2022}.
Extending these experiments to multiple impurities with all spin and
channel species propagating between the impurities is a challenge.
Testing if more realistic setups, in which only some of the species
connect the impurities while the remainder are local to each impurity,
also support non-Abelian anyons, and what physical observables probe
their fusion space, is quite straightforward for the method presented,
and is left for future work. Note that due to the absence of a (topological)
gap, we expect information encoded in the fusion space to decohere
as a power law of $T/T_{K}$, in contrast to the exponential suppression
in the presence of a gap. Still, based on the success of Refs.\,\citep{iftikharTwochannelKondoEffect2015,iftikharTunableQuantumCriticality2018},
the path to observing non-Abelian anyons might be shorter in these
systems.
\begin{acknowledgments}
\textit{Acknowledgments.---} We would like to thank J. von Delft,
A. Weichselbaum, S.-S. Lee and K. Shtengel for fruitful discussions.
Numerical simulations were performed using the \textsc{QSpace} tensor-network
library and accompanying code \citep{weichselbaumTensorNetworksNumerical2012,weichselbaumNonabelianSymmetriesTensor2012,weichselbaumXsymbolsNonAbelianSymmetries2020,leeAdaptiveBroadeningImprove2016}.
E.S. was supported by the Synergy funding for Project No. 941541,
ARO (W911NF-20-1-0013), the US-Israel Binational Science Foundation
(BSF) Grant No. 2016255, and the Israel Science Foundation (ISF) Grant
No. 154/19. M.G. was supported by the ISF and the Directorate for
Defense Research and Development (DDR\&D) Grant No. 3427/21 and by
the BSF Grant No. 2020072.
\end{acknowledgments}

\bibliography{../../../bib,foot}

\onecolumngrid

\newpage{}

\include{ChiralKondoZL_Supplemental}

\end{document}

%% file: ChiralKondoZL_Supplemental.tex
\begin{center}
\textbf{\large{}Supplemental Material for }\\
\textbf{\large{}``Manipulating Non-Abelian Anyons in a Chiral Multichannel
Kondo Model''}{\large\par}
\par\end{center}

\setcounter{equation}{0}
\setcounter{figure}{0}

\renewcommand{\theequation}{S\arabic{equation}}
\renewcommand{\thefigure}{S\arabic{figure}}
\renewcommand{\thetable}{S\Roman{table}}

This supplemental consists of four sections. Sec.\,\ref{sec:CFT}
shortly reviews relevant results and implications of the multi-impurity
CFT ansatz of \citet{lopesAnyonsMultichannelKondo2020}, and specifically
the resulting finite-size spectrum. In Sec.\,\ref{sec:Finite-Size-Spectrum}
we analyze the NRG low-energy spectrum and demonstrate its agreement
with the CFT ansatz finite-size spectrum. Sec.\,\ref{sec:NRG-Details}
contains NRG implementation instructions followed by a discussion
of the model symmetries and how to exploit them. Finally, in Sec.\,\ref{sec:Parameters}
we discuss the model and NRG parameters used throughout. 

\section{Implications of the CFT Ansatz\label{sec:CFT}}

\citet{affleckKondoEffectConformal1991} conjectured (and then demonstrated
by comparing to both analytical and numerical solutions) that the
low-energy physics of the single spin-$s$ impurity $k$-channel Kondo
effect is captured by a boundary CFT in which a single $\mathrm{SU}\!\left(2\right)_{k}$
``anyon'' with ``topological'' charge $s$ is fused to the primary
fields of $k$ channels of free fermions. Here ``anyons'' of charge
$s$ are in one to one correspondence with the primary field of the
$\mathrm{SU}\!\left(2\right)_{k}$ Wess-Zumino-Witten theory, labeled
by $s=0,\frac{1}{2},1,...,\frac{k}{2}$, and satisfying the fusion
rules given in Eq.\,(\ref{eq:Fusion}) below. Generalizing upon this,
\citet{lopesAnyonsMultichannelKondo2020} conjectured that for $M$
spin-$s$ impurities, and assuming the $k$ channels are chiral, the
low-energy physics is captured by fusing $M$ such $\mathrm{SU}\!\left(2\right)_{k}$
anyons, each with charge $s$, to the primary fields of $k$ channels
of free fermions. Here we will outline some of the consequences of
this conjecture for two examples: $M$ spin-$\frac{1}{2}$ impurities
with $k=2$ or $k=3$ channels. 
\begin{table}[b]
\begin{centering}
\begin{tabular}{|c|c|}
\hline 
$\begin{array}{lcc}
\qquad\text{\ensuremath{\mathrm{SU\!\left(2\right)_{2}}} Anyons} & \!\!\!\!\!\!\!\!\longrightarrow\ \  & \text{Ising Anyons (Majoranas)}\\
\begin{array}{l}
0\times s=s\ ;\ s\in\left\{ 0,\tfrac{1}{2},1\right\} \\
\!\tfrac{1}{2}\times\tfrac{1}{2}=0+1\\
\!\tfrac{1}{2}\times1=\tfrac{1}{2}\\
1\times1=0
\end{array} & \!\!\!\!\!\!\!\!\longrightarrow\ \  & \begin{array}{l}
\mathbf{1}\times\alpha=\alpha\ ;\ \alpha\in\left\{ \mathbf{1},\sigma,\psi\right\} \\
\sigma\times\sigma=\mathbf{1}+\sigma\\
\sigma\times\psi=\sigma\\
\psi\times\psi=\mathbf{1}
\end{array}\\
 & \vspace{-1em}
\end{array}$ & $\begin{array}{lcc}
\quad\qquad\text{\ensuremath{\mathrm{SU\!\left(2\right)_{3}}} Anyons} & \!\!\!\!\!\!\!\!\!\!\!\!\longrightarrow\ \  & \text{Fibonacci Anyons}\\
\begin{array}{l}
\!\!\begin{array}{l}
0\times s=s\\
\!\tfrac{3}{2}\times s=\tfrac{3}{2}-s
\end{array};\ s\in\left\{ 0,\tfrac{1}{2},1,\tfrac{3}{2}\right\} \\
\!\tfrac{1}{2}\times\tfrac{1}{2}=1\times1=0+1\\
\!\tfrac{1}{2}\times1=\tfrac{1}{2}+\tfrac{3}{2}
\end{array} & \!\!\!\!\!\!\!\!\!\!\!\!\longrightarrow\ \  & \begin{array}{l}
\mathbf{1}\times\alpha=\alpha\ ;\ \alpha\in\left\{ \mathbf{1},\tau\right\} \\
\,\\
\tau\times\tau=\mathbf{1}+\tau
\end{array}\\
 & \vspace{-1em}
\end{array}$\tabularnewline
\hline 
\end{tabular}
\par\end{centering}
\caption{Fusion rules for different anyonic models. For $\mathrm{SU}\!\left(2\right)_{2}$
anyons we can associate the topological charges $\left\{ 0,\frac{1}{2},1\right\} $
with the vacuum $\mathbf{1}$, a Majorana fermion $\sigma$, and an
occupied Dirac fermion $\psi$, respectively, arriving at the fusion
rules of Ising anyons. For $\mathrm{SU}\!\left(2\right)_{3}$ anyons
we associate both $0$ and $\frac{3}{2}$ with the vacuum $\mathbf{1}$,
and both $\frac{1}{2}$ and $1$ with the $\tau$ anyon, arriving
at the fusion rules of Fibonacci anyons. \label{tab:Fusion-rules}}
\end{table}

The fusion rule of two $\mathrm{SU}\!\left(2\right)_{k}$ anyons with
topological charges $s_{1},s_{2}\leq\frac{k}{2}$ is given by
\begin{equation}
s_{1}\times s_{2}=\left|s_{1}-s_{2}\right|+\left(\left|s_{1}-s_{2}\right|+1\right)+...+\min\left(s_{1}+s_{2},k-s_{1}-s_{2}\right),\label{eq:Fusion}
\end{equation}
which for $k\to\infty$ is simply the standard $\mathrm{SU}\!\left(2\right)$
fusion rule, i.e., the angular momentum addition rule. Specifically
for $k=2$ and $k=3$ we get the fusion rules in Table \ref{tab:Fusion-rules},
which can be identified with those of Ising and Fibonacci anyons,
respectively (see caption). Fusing $M$ charge-$\frac{1}{2}$ anyons
according to these rules we get
\begin{align}
\mathrm{SU\!\left(2\right)_{2}}:\quad\overset{M}{\overbrace{\tfrac{1}{2}\times\tfrac{1}{2}\times...\times\tfrac{1}{2}}} & =\Biggl\{\begin{array}{ll}
\overset{2^{\left(M-1\right)/2}}{\overbrace{\tfrac{1}{2}+\tfrac{1}{2}+...+\tfrac{1}{2}}} & \text{odd }M\\
\underset{2^{M/2-1}}{\underbrace{0+...+0}}+\underset{2^{M/2-1}}{\underbrace{1+...+1}} & \text{even }M
\end{array},\label{eq:SU2-2-fusion}\\
\mathrm{SU\!\left(2\right)_{3}:\quad\overset{M}{\overbrace{\tfrac{1}{2}\times\tfrac{1}{2}\times...\times\tfrac{1}{2}}}} & =\Biggl\{\begin{array}{ll}
\overset{F_{M}}{\overbrace{\tfrac{1}{2}+...+\tfrac{1}{2}}}+\overset{F_{M-1}}{\overbrace{\tfrac{3}{2}+...+\tfrac{3}{2}}} & \text{odd }M\\
\underset{F_{M-1}}{\underbrace{0+...+0}}+\underset{F_{M}}{\underbrace{1+...+1}} & \text{even }M
\end{array},\label{eq:SU2-3-fusion}
\end{align}
where $F_{M}$ are the elements of the Fibonacci sequence $F_{0}=0,F_{1}=1,F_{M>1}=F_{M-1}+F_{M-2}$.
Projecting onto any specific fusion outcome, we get a (fusion) Hilbert
space which grows exponentially with the number of anyons. For $k=2$,
adding two anyons doubles the size of the fusion space, and so a single
anyon carries a quantum dimension of $\sqrt{2}$. For $k=3$ we see
that it make sense to associate both $\frac{1}{2}$ and $1$ with
the $\tau$ anyon and $0$ and $\frac{3}{2}$ with the vacuum (as
in Table \ref{tab:Fusion-rules}), and then adding a single $\tau$
anyon enlarges the dimension of the fusion space by $F_{M}/F_{M-1}$,
which in the large $M$ limit gives the golden ratio $\frac{1+\sqrt{5}}{2}$,
i.e., the quantum dimension associated with Fibonacci anyons.

Returning to the multi-impurity fusion ansatz we note that due to
the associativity of fusion, we can first fuse the $M$ charge-$\frac{1}{2}$
anyons to each other, arriving at the anyonic fusion space of Eq.\,(\ref{eq:SU2-2-fusion})
or (\ref{eq:SU2-3-fusion}), and then fuse this space onto the free
fermions. In order to calculate the spectrum, we simply need to solve
for single spin-$s=\frac{1}{2},1,...,\frac{k}{2}$ impurities coupled
to $k$ channels, and add the degeneracies according to the fusion
outcomes by hand. In Sec.\,\ref{sec:Finite-Size-Spectrum} we will
demonstrate that the low-energy spectrum obtained from NRG (for two
impurities with two and three channels, and for three impurities with
two channels) exactly matches this structure.

\section{Finite-Size Spectrum\label{sec:Finite-Size-Spectrum}}

\begin{figure*}

\begin{minipage}[t]{0.48\columnwidth}%
\phantomsubfloat{\label{fig:Spectrum-2I2CK}}\phantomsubfloat{\label{fig:Spectrum-2I3CK}}\phantomsubfloat{\label{fig:Spectrum-3I2CK}}

\begin{figure}[H]
\caption{NRG low-energy spectrum at the chiral point, i.e., critical hopping
$t_{c}^{\prime}$ (gray), and away from it (red and blue). The chiral
spectrum agrees with the $\mathrm{SU}\!\left(2\right)_{k}$ CFT finite-size
spectrum for $k$ channels and different number of impurities (anyons).
(a,b) For two spin-$\frac{1}{2}$ impurities it is given by overlaying
the spectrum of a spin-0 (blue) and spin-1 (red) $k$-channel Kondo
spectrum, while (c) for three impurities it is given by two identical
copies of the spin-$\frac{1}{2}$ $k$-channel Kondo spectrum (red).
Observe that the degeneracies (black text), which for the chiral spectrum
(gray) are partitioned according to the expectation values of the
spin correlator, indeed match those of the corresponding fusion channel.
NRG parameters $\Lambda=2$ for (a,b) and $\Lambda=3$ for (c), $N_{K}=4000$,
and model parameters $J=\Gamma=8D$ are chosen for optimal numerical
convergence, as discussed in Sec.\,\ref{sec:Parameters}. \label{fig:Spectrum}}
\end{figure}
\end{minipage}~~~~~%
\begin{minipage}[t]{0.48\columnwidth}%
\begin{figure}[H]
\includegraphics[width=1\columnwidth]{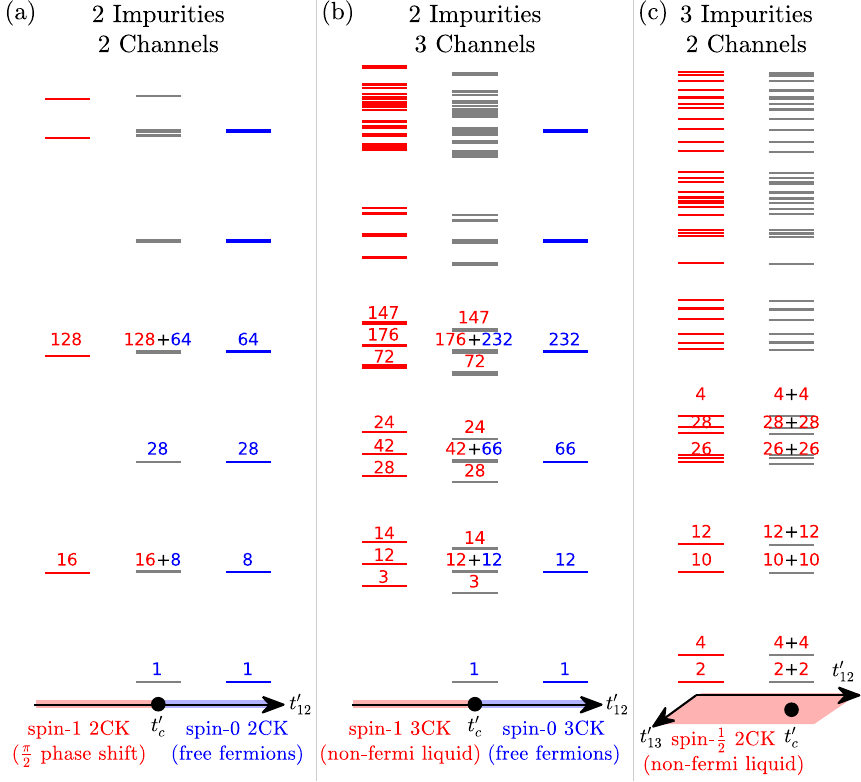}
\end{figure}
\end{minipage}

\end{figure*}

In this section we study the effective Hamiltonian at the low-energy
fixed point, as obtained by NRG. As the system is gapless, in the
thermodynamic limit it has a continuous spectrum, but at any given
NRG iteration we can interpret the spectrum as that of a finite system,
in which case we do have discrete levels. One needs to choose the
boundary conditions of this finite system, and they dictate whether
the bath has an even or odd number of single-particle modes. For conciseness,
we will restrict our discussion to an even number of such modes, corresponding
to even NRG iterations, and to half filling, so that there is no single-particle
level at the Fermi energy. As we exploit the global symmetries of
the model as part of the NRG procedure, we retain the information
regarding the quantum numbers, i.e., symmetry labels, of each one
of the many-body levels in the low-energy spectrum. We will now demonstrate
that the obtained spectrum is in perfect agreement with the finite-size
spectrum of the boundary CFT ansatz of Ref.\,\citep{lopesAnyonsMultichannelKondo2020},
which is obtained by $\mathrm{SU}\!\left(2\right)_{k}$ fusion rules
as outlined in Sec.\,\ref{sec:CFT}.

\subsection{Two Impurities}

We start with two impurities and two or three channels. By setting
$t_{12}^{\prime}$, the hopping amplitude between the dangling sites,
to be above (below) the critical (chiral) value $t_{c}^{\prime}$,
we get the low-energy fixed-point spectra of a single spin-0 (spin-1)
impurity Kondo effect, as plotted in blue (red) in Figs.\,\ref{fig:Spectrum-2I2CK}
and \ref{fig:Spectrum-2I3CK}, for two and three channels respectively.
These can be calculated, e.g., via boundary CFT \citep{affleckKondoEffectConformal1991}
by fusing an $\mathrm{SU}\!\left(2\right)_{k}$ anyon of charge-$s$
to the primary fields of ($k$-channel) free fermions. For $s=0$
this simply leaves us with $k$ channels of free fermions, which can
easily be verified. For spin-1 and two channels, each channel contributes
a single spin-$\frac{1}{2}$ fermion, thus fully screening the impurity,
and resulting in a $\frac{\pi}{2}$ phase shift, i.e., free fermions
with an odd number of single particle modes (assuming we started with
an even number of modes). For spin-1 and three channels we get an
over-screened effect, which results in a non-fermi liquid. Tuning
to the critical point, we find that there is no hybridization between
the energy levels of the two spectra, which are simply overlayed,
as demonstrated in gray in Figs.\,\ref{fig:Spectrum-2I2CK} and \ref{fig:Spectrum-2I3CK}.
This is exactly the picture one expects when fusing two charge-$\frac{1}{2}$
anyons to a $k$-channel bath: 
\[
\tfrac{1}{2}\times\tfrac{1}{2}\times\mathrm{Bath}=\left(0+1\right)\times\mathrm{Bath}=0\times\mathrm{Bath}+1\times\mathrm{Bath},
\]
or in words, we first fuse the two anyons to each other, resulting
in an anyonic charge of either 0 or 1, which is then separately fused
to the bath, arriving at the combined spectrum. Thus, we understand
that levels associated with the different spectra belong to different
fusion channels, which explains the absence of hybridization. For
two impurities, most levels in the combined spectrum are associated
either with only one of the fused charges, as is, e.g., the case for
the many-body ground state. However, we do have levels which are associated
with both channels, e.g., the first (second) excited level for $k=2$
($k=3$) channels, in which case we observe anyonic degeneracy.

As discussed in the main text, projecting the interimpurity spin correlator
$\mathbf{S}_{1}\!\cdot\mathbf{S}_{2}$ onto the low-energy subspace
results in an operator which commutes with the low-energy Hamiltonian,
and so we can mutually diagonalize them. The low-energy projected
$\mathbf{S}_{1}\!\cdot\mathbf{S}_{2}$ has only two distinct eigenvalues,
one negative (singletlike) and one positive (tripletlike). Calculating
the relevant eigenvalue for each level in the spectra in Figs.\,\ref{fig:Spectrum-2I2CK}
and \ref{fig:Spectrum-2I3CK}, we indicate in blue (red) the degeneracies
of the levels with the negative (positive) eigenvalue, and find that
they exactly match those associated with the spin-0 (spin-1) spectrum.
We thus establish that $\mathbf{S}_{1}\!\cdot\mathbf{S}_{2}$ indeed
measures the fusion outcome. Note that away from the critical point,
only one fusion outcome survives, and indeed the low-energy projected
$\mathbf{S}_{1}\!\cdot\mathbf{S}_{2}$ is proportional to the identity
in these cases.

We point out that due to the NRG discretization, we expect (and observe)
corrections (with respect to the CFT spectra), i.e., splitting and
shifting of energy levels, which become more pronounced at higher
energy levels. However, these corrections affect the spectra in a
systematic manner without breaking the correspondence between the
spectrum at the critical point and the spectra of the two phase away
from it. Still, for a clear comparison we choose the model parameters
as discussed in Sec.\,\ref{sec:Parameters}.

\subsection{Three Impurities}

Going to three impurities and two channels, we have three imaginary
hopping amplitudes which we need to tune. The nearest-neighbor terms
$t_{12}^{\prime}$ and $t_{23}^{\prime}$ are equal by (time reversal
+ inversion) symmetry, while the next-nearest-neighbor term $t_{13}^{\prime}$
can have a different value. In the absence of the artificial cutoff
$D$ all three should be equal to $\Gamma$, which implies a single
critical point in a two-dimensional parameter space. As in the two-impurity
case, we expect any deviation from this critical point to send us
to one of the two fusion outcomes. However, in the case of three impurities
and two channels we have 
\[
\tfrac{1}{2}\times\tfrac{1}{2}\times\tfrac{1}{2}\times\mathrm{Bath}=\left(\tfrac{1}{2}+\tfrac{1}{2}\right)\times\mathrm{Bath}=\tfrac{1}{2}\times\mathrm{Bath}+\tfrac{1}{2}\times\mathrm{Bath},
\]
so that the two fusion outcomes correspond to identical phases, i.e.,
a single-impurity 2CK low-energy spectra, as shown in red in Fig.\,\ref{fig:Spectrum-3I2CK}.
Still, they do differ in their fusion path, e.g., in the fusion outcome
of the first two impurities, and indeed we find that the interimpurity
spin correlator $\mathbf{S}_{1}\!\cdot\mathbf{S}_{2}$ has two distinct
regimes (the same holds for $\mathbf{S}_{1}\!\cdot\mathbf{S}_{3}$,
and trivially by symmetry for $\mathbf{S}_{2}\!\cdot\mathbf{S}_{3}$).
Generally the crossover between these two regimes is smooth, with
only a single point of discontinuity. Tuning exactly to this point,
we get the spectrum of three spin-$\frac{1}{2}$ impurities fused
to two-channel bath, as shown in gray in Fig.\,\ref{fig:Spectrum-3I2CK}.
We can again count the levels associated with each of the eigenvalues
of $\mathbf{S}_{1}\!\cdot\mathbf{S}_{2}$ (or $\mathbf{S}_{2}\!\cdot\mathbf{S}_{3}$
or $\mathbf{S}_{1}\!\cdot\mathbf{S}_{3}$) and indeed we find that
for each level with the positive eigenvalue we have one with the negative
eigenvalue.

\section{NRG Implementation Details\label{sec:NRG-Details}}

\begin{figure}
\begin{centering}
\includegraphics[width=0.9\columnwidth]{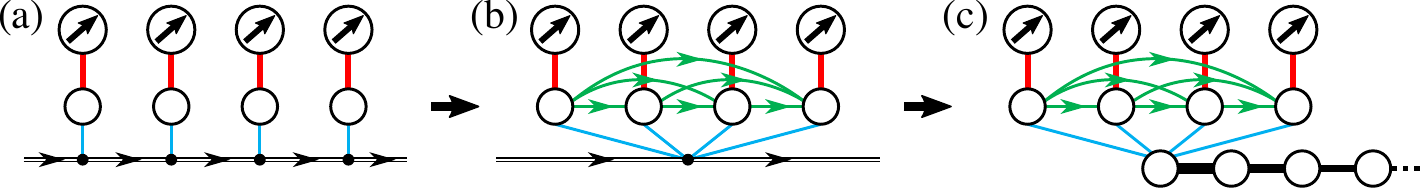}
\par\end{centering}
\begin{centering}
\phantomsubfloat{\label{fig:Sup-Dangling-Model}}\phantomsubfloat{\label{fig:Sup-Effective-Model}}\phantomsubfloat{\label{fig:Sup-Wilson-Chain-Model}}\vspace{-2.5em}
\par\end{centering}
\caption{Mapping of (a) the original dangling-site model to (b) the effective
model in the limit of very close impurities, and then (c) after replacing
the trivial bath by a Wilson chain as in Eq.\,(\ref{eq:H-disc}).\label{fig:Wilson-chain}}

\end{figure}

Here we will outline the technical details of applying NRG to our
effective model. We assume familiarity with the NRG jargon, and stop
short of reviewing the iterative diagonalization, which can be carried
out by a traditional \citep{bullaNumericalRenormalizationGroup2008}
or contemporary (tensor-network based) \citep{weichselbaumTensorNetworksNumerical2012}
implementation. We will start by discussing the mapping to a Wilson
chain (although this is standard procedure), and follow with a discussion
about exploiting symmetries of the problem.

\subsection{Mapping to a Wilson Chain}

Our starting point is the Hamiltonian in Eq.\,(\ref{eq:H-total}),
depicted in Fig.\,\ref{fig:Sup-Dangling-Model}. After replacing
$H_{\mathrm{dang}}$ with $H_{\mathrm{dang}}^{\mathrm{eff}}$ of Eq.\,(\ref{eq:H-effective}),
we arrive at the effective Hamiltonian depicted in Fig.\,\ref{fig:Sup-Effective-Model},
which describes a large effective impurity consisting of the original
impurity spins together with the dangling sites (and the imaginary
couplings between them, which encode the chirality), coupled to a
trivial bath given by $H_{\mathrm{chiral}}$ of Eq.\,(\ref{eq:H-chiral}).
It is straightforward to diagonalize the bath by going to $k$-space
\begin{equation}
H_{\mathrm{chiral}}=\sum_{\alpha\sigma}\int_{-\infty}^{\infty}\frac{dk}{2\pi}v_{F}kc_{k\alpha\sigma}^{\dagger}c_{k\alpha\sigma},\quad c_{k\alpha\sigma}=\int_{-\infty}^{\infty}dxe^{-ikx}\psi_{\alpha\sigma}\left(x\right).\label{eq:H-chiral-diag}
\end{equation}
We then introduce a sharp high-energy cutoff $v_{F}k_{\max}=D$ and
logarithmically discretize \citep{bullaNumericalRenormalizationGroup2008,campoAlternativeDiscretizationNumerical2005,zitkoAdaptiveLogarithmicDiscretization2009}
Eq.\,(\ref{eq:H-chiral-diag}). Tridiagonalizing the full (effective)
Hamiltonian, we arrive at
\begin{align}
H_{\mathrm{disc}} & =J\sum_{m}\mathbf{S}_{m}\cdot\mathbf{s}_{m}+\sum_{\alpha\sigma}\sum_{m>m^{\prime}}it_{mm^{\prime}}^{\prime}\left(d_{m\alpha\sigma}^{\dagger}d_{m^{\prime}\alpha\sigma}-d_{m^{\prime}\alpha\sigma}^{\dagger}d_{m\alpha\sigma}\right)\nonumber \\
 & +\sqrt{M}t_{0}\sum_{m\alpha\sigma}\left(d_{m\alpha\sigma}^{\dagger}c_{0\alpha\sigma}+c_{0\alpha\sigma}^{\dagger}d_{m\alpha\sigma}\right)+\sum_{\alpha\sigma}\sum_{n=0}^{N}t_{n}\left(c_{n\alpha\sigma}^{\dagger}c_{\left(n+1\right)\alpha\sigma}+c_{\left(n+1\right)\alpha\sigma}^{\dagger}c_{n\alpha\sigma}\right),\label{eq:H-disc}
\end{align}
with $t_{0}=\sqrt{\frac{2}{\pi}\Gamma D}$ and $t_{n}\sim D\Lambda^{-n/2}$
the exponentially decaying hopping amplitudes along the Wilson chain,
which is of some finite length $N$. Thus, the effective impurity
is now coupled to the first site of a standard Wilson chain, as depicted
in Fig.\,\ref{fig:Sup-Wilson-Chain-Model}. We then proceed to iteratively
diagonalize this chain, adding one chain site at a time, diagonalizing
the new Hamiltonian, and truncating to a fixed number $N_{K}$ of
low-energy states. For each iteration we thus have an effective low-energy
spectrum which we can analyze (see Sec.\,\ref{sec:Finite-Size-Spectrum}),
and from which can extract thermodynamic quantities, e.g., the impurity
entropy, associated with a temperature $T\sim D\Lambda^{-n/2}$. We
stop the iterative diagonalization, i.e., choose $N$, once we see
that all quantities have converged.

\subsection{Exploiting Symmetries}

In order to reduce the computational cost of the iterative diagonalization
procedure, we exploit global symmetries, which is quite natural when
formulating NRG in terms of a tensor-network algorithm. In the case
of Abelian symmetries, the tensors (or matrices) break down into a
block structure, so that each block is diagonalized separately, thus
allowing larger tensors, or a larger number of kept states $N_{K}$
in each iteration (for the same computational cost). The significant
advantage comes from exploiting non-Abelian symmetries, in which case
each block can be decomposed into an outer product of ``actual information''
(akin to reduced matrix elements in the Wigner-Eckart theorem \citep{sakuraiModernQuantumMechanics2017b})
and ``symmetry structure'', with the later encoded in (generalized)
Clebsch-Gordan coefficients \citep{weichselbaumNonabelianSymmetriesTensor2012}.
We then need to only diagonalize the ``actual information'' part
which significantly reduces the size of each block.

The symmetries typically associated with the multichannel Kondo problem
are $\mathrm{SU\!\left(2\right)}$ spin symmetry (in the absence of
a magnetic field), as well as $\mathrm{U\!\left(1\right)}$ charge
(particle number) and $\mathrm{SU\!\left(k\right)}$ channel symmetries,
or compactly, $\mathrm{U\!\left(1\right)_{charge}\times SU\!\left(k\right)_{channel}\times SU\!\left(2\right)_{spin}}$
symmetry. The CFT ansatz is formulated in terms of the these symmetries,
and one can easily verify that both the original and effective discrete
multi-impurity Hamiltonians in Eqs.\,(\ref{eq:H-total}) and (\ref{eq:H-disc}),
respectively, conserve these symmetries. We also consider the system
in a particle-hole symmetric regime, in which case the charge and
channel symmetries are elevated to an $\mathrm{Sp\!\left(2k\right)}$
symmetry \citep{weichselbaumNonabelianSymmetriesTensor2012}, so that
the full symmetry of the model is actually $\mathrm{Sp\!\left(2k\right)_{\text{charge-channe}l}\times SU\!\left(2\right)_{spin}}$.
This is a direct generalization of the more widely known $\mathrm{SU\!\left(2\right)_{charge}\times SU\!\left(2\right)_{spin}}$
symmetry in the single-channel case. We can actually stick to our
intuition for the single-channel case, and resolve all arising complications
at that level, with specific technical challenges for exploiting the
$\mathrm{Sp\!\left(2k\right)}$ symmetry taken care at the level of
the tensor-network library \citep{weichselbaumNonabelianSymmetriesTensor2012,weichselbaumXsymbolsNonAbelianSymmetries2020}.

We start by formulating the particle-hole transformation under which
the Hamiltonian is invariant
\begin{equation}
\psi_{\alpha\sigma}\left(x\right)\leftrightarrow\psi_{\alpha\sigma}^{\dagger}\left(x\right)\ \Leftrightarrow\ c_{k\alpha\sigma}\leftrightarrow c_{-k\alpha\sigma}^{\dagger}\ \Leftrightarrow\ c_{n\alpha\sigma}\leftrightarrow\left(-1\right)^{n}c_{n\alpha\sigma}^{\dagger}\quad;\quad d_{m\alpha\sigma}\leftrightarrow-d_{m\alpha\sigma}^{\dagger}
\end{equation}
noting that this, together with the $\mathrm{U\!\left(1\right)_{charge}}$
symmetry suffices in to claim $\mathrm{SU\!\left(2\right)_{charge}}$
symmetry in the single-channel case (and $\mathrm{Sp\!\left(2k\right)}$
symmetry in the $\mathrm{SU\!\left(k\right)_{channel}}$ symmetric
case). Observe that the dangling-site operators flip sign under the
transformation, as do the odd sites in the Wilson chain. It is usually
argued that under $\mathrm{SU\!\left(2\right)_{charge}}$ symmetry
we can partition our tight-binding model into a bipartite graph (in
our case even chain sites on the one side and odd chain sites together
with the dangling sites on the other), allowing only (purely real)
couplings which cross the partition. However, this restriction assume
the system is invariant under time reversal $t\to-t$, while a chiral
system is not, and is only invariant under time reversal together
with inversion, $t\to-t,x\to-x$. In this case purely imaginary couplings
between sites on the same side of the partition also respect the symmetry,
e.g., as in the case of the dangling sites. On a technical level we
note that if one can only implement (real) terms which cross the partition,
e.g., $\sum_{\alpha\sigma}\left(d_{m\alpha\sigma}^{\dagger}c_{0\alpha\sigma}+c_{0\alpha\sigma}^{\dagger}d_{m\alpha\sigma}\right)$,
one can still obtain the imaginary terms by calculating commutators
\begin{equation}
\sum_{\alpha\sigma}i\left(d_{m\alpha\sigma}^{\dagger}d_{m^{\prime}\alpha\sigma}-d_{m^{\prime}\alpha\sigma}^{\dagger}d_{m\alpha\sigma}\right)=i\left[\sum_{\alpha\sigma}\left(d_{m\alpha\sigma}^{\dagger}c_{0\alpha\sigma}+c_{0\alpha\sigma}^{\dagger}d_{m\alpha\sigma}\right),\sum_{\alpha\sigma}\left(d_{m^{\prime}\alpha\sigma}^{\dagger}c_{0\alpha\sigma}+c_{0\alpha\sigma}^{\dagger}d_{m^{\prime}\alpha\sigma}\right)\right].
\end{equation}

\section{Model and NRG Parameters\label{sec:Parameters}}

We start by commenting on the main tunable NRG parameters. The logarithmic
discretization parameter $\Lambda$ defines the decay rate of the
hopping amplitudes along the Wilson chain. While NRG is formally exact
in the limit $\Lambda\to1$, in order to justify the iterative diagonalization
we require $\Lambda>1$, and it is common practice to take $\Lambda\lesssim3$,
which typically suffices in most cases. We use a dynamical truncation
scheme, specified by a rescaled truncation energy $E\left(n\right)=E_{K}\Lambda^{-n/2}$,
and strive in each iteration to keep the $N_{E\left(n\right)}$ multiplets
with energies below it. We also introduce a maximal number of kept
multiplets $N_{K}$, dictated by what is computationally tractable.
The specific choice of $E_{K}$ is not very important (we use $10\frac{\Lambda+1}{2}D$),
but for a given choice, $N_{E\left(n\right)}$ does serve as a measure
for the number of required kept multiplets (in a given iteration)
in order to keep numerical errors under control. Thus, satisfying
the limit $N_{K}$ is a good indication the calculation might not
be fully converged. Note that for a fixed $E_{K}$, taking larger
$\Lambda$ implies a smaller $N_{E\left(n\right)}$, and unless stated
otherwise we take $\Lambda=3$. We crank the maximal limit $N_{K}$
up to 12,000 kept multiplets, which due to the use of symmetries as
explained in Sec.\,\ref{sec:NRG-Details}, corresponds to 600,000
(4,800,000) kept states for the $k{=}2$ ($k{=}3$) channels and $M{=}3$
($M{=}2$) impurities calculations.

The main numerical challenge we encounter is the exponential scaling
of the size of the Hilbert space of the effective impurity, both with
the number of channels $k$ and the number of impurities $M$. It
contains $M$ impurity spins and $k\times M$ spinfull fermionic modes
(of the dangling sites), so that even before coupling to the Wilson
chain we start with a Hilbert space of dimension $2^{\left(2k+1\right)M}$.
Introducing the first Wilson chain site multiplies the size of the
Hilbert space by $2^{2k}$. For $k{=}2$ ($k{=}3$) channels and $M{=}3$
($M{=}2$) impurities this leads to a $2^{19}\times2^{19}$ ($2^{20}\times2^{20}$)
matrix that needs to be diagonalized, which is beyond what is tractable
in the absence of symmetries. Even after exploiting all the symmetries
of the model, as discussed in Sec.\,\ref{sec:NRG-Details}, we start
with hundreds of multiplets for the effective impurity, so that by
the second NRG iteration we surpass $N_{K}$ and are forced to start
truncating to low-energy states.

We now turn to discuss the optimal range for the model parameters.
First of all, we would ideally like to take the artificial cutoff
$D$ to be larger than all other energy scales in order to mitigate
its effect, i.e., the impairing of chirality which needs to be corrected
by tuning the dangling-site imaginary hopping amplitudes $t_{mm}^{\prime}$,
as demonstrated in Fig.\,\ref{fig:chiral-hopping}. We would ideally
also like to take $J<\Gamma$, in which case the Kondo temperature
is given by $T_{K}=\Gamma e^{-\pi\Gamma/J}$ and is thus exponentially
smaller than the bare energy scales, so that we have a nice separation
between different physical regimes. However, although NRG is explicitly
constructed to treat such energy-scale separation, some parameter
regimes require more computational resources than others, making it
desirable to steer away from the ideal limit. Let us examine how this
turns out in practice, starting from the impurity entropy, which only
depends on the gross features of the energy spectrum, going on to
the $F$ matrices, which depend on the wavefunctions, and finally
studying the full energy spectrum.

\subsection{Impurity Entropy}

\begin{figure}
\begin{centering}
\includegraphics[width=1\columnwidth]{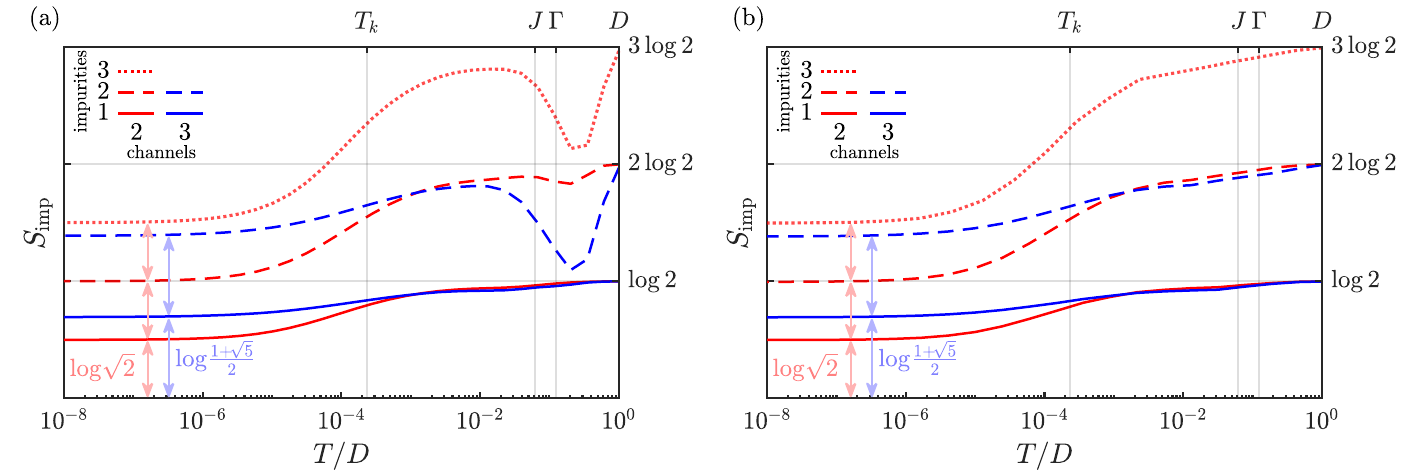}
\par\end{centering}
\begin{centering}
\phantomsubfloat{\label{fig:Impurity-entropy-A}}\phantomsubfloat{\label{fig:Impurity-entropy-B}}\vspace{-2.5em}
\par\end{centering}
\caption{Impurity entropy per impurity for two and three channels (red and
blue, respectively) and up to three impurities, with $2J=\Gamma=D/8$,
and different choices of NRG parameters: (a) $\Lambda=3,N_{K}\protect\leq12000$
converges to the correct low-energy fixed point, but displays truncation
artifacts at high temperatures. (b) Parameters as in Fig.\,\ref{fig:Impurity-entropy}
in the main text: $\Lambda=6$ ($\Lambda=8$) for 1,2 (3) impurities,
and $N_{K}\protect\leq12000$, which eliminate the artifacts.}
\end{figure}

If we take $D\gg J,\Gamma$, then in the first few NRG iterations
the Wilson chain couplings are larger than the bare effective-impurity
energy scales $J$ and $\Gamma$, and we cannot discriminate between
different impurity many-body states. Thus, the we need to keep many
states, i.e., we have a large $N_{E\left(n\right)}$, but due to the
colossal size of the effective impurity, for reasonable $\Lambda=3$
we quickly saturate the $N_{K}$ limit and are forced to discard some
required states. As a result we lose accuracy, which leads to an artificial
drop in the impurity entropy at high temperatures in Fig.\,\ref{fig:Impurity-entropy-A}.
Once we go below the bare energy scales, we start seeing the separation
between the impurity states, and those of higher energy no longer
mix with those of lower energy, so that they can be discarded. Thus,
in this regime the number of kept multiplets $N_{E\left(n\right)}$
required for the same accuracy is smaller, and falls beneath the threshold
$N_{K}$. Going below $T_{K}$, the required $N_{E\left(n\right)}$
shrinks even further. The errors accumulated from the uncontrolled
truncations at the beginning of the chain enter the RG Hamiltonian
in the form of irrelevant corrections, and so once we can keep enough
multiplets, the RG flow corrects itself. As a result the impurity
entropy, which is explicitly extracted from the RG Hamiltonian, also
corrects itself, and follows the universal curve of the single-impurity
entropy, multiplied by the number of impurities. We can improve the
situation at high energies without compromising the low-energy behavior
by relying on the observation that the impurity entropy is not very
sensitive to $\Lambda$. Thus, taking relatively large $\Lambda=6,8$
we are able to reduce the number of required kept multiplets to what
is tractable even above the bare scales, hence eliminate the artifacts
while maintaining an overall correct behavior, as demonstrated in
Fig.\,\ref{fig:Impurity-entropy-B}.

\subsection{$F$ Matrices}

Turning to the eigenfunctions, they are more sensitive than the spectrum,
and hence may suffer uncorrectable errors in the early regime. This
affects observables which depend on them, such as the interimpurity
spin correlators $\mathbf{S}_{m}\!\cdot\mathbf{S}_{m^{\prime}}$ and
the extracted $F$ matrices. We thus wish to mitigate these errors
by taking the shortest route to the low-energy anyonic fixed point.
This is achieved by taking $J\gtrsim\Gamma$ so that $T_{K}$ is also
of the same order. We take all three to be smaller than the artificial
cutoff $D$, so that it does not lead to significant errors, but not
much smaller so that we have a small region in which we have to carry
out uncontrolled truncations. In what follows we will quantitatively
demonstrate this.

\begin{figure*}

\begin{minipage}[t]{0.54\columnwidth}%
\phantomsubfloat{\label{fig:chiral-hopping}}\phantomsubfloat{\label{fig:EV-ratio}}\vspace{-2.8em}

\includegraphics[width=1\columnwidth]{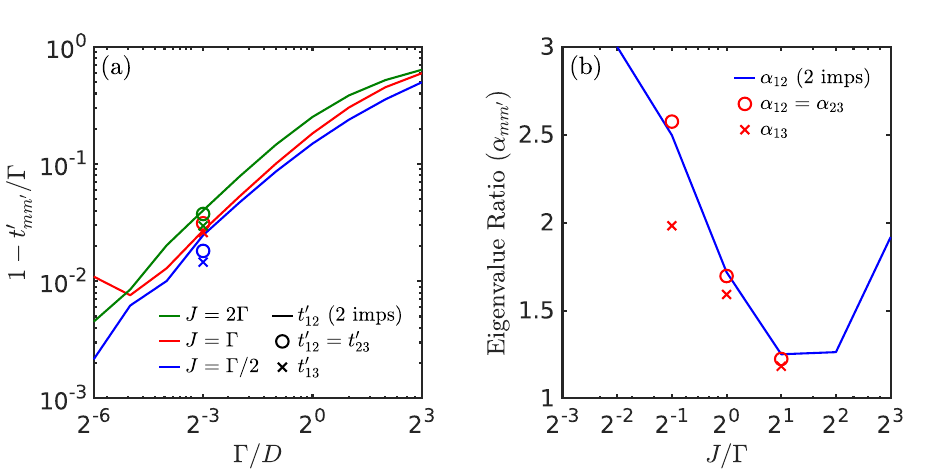}\vspace{-2em}

\begin{figure}[H]
\caption{(a) Convergence of the critical chiral hopping amplitudes $t_{mm}^{\prime}$
to $\Gamma$ in the limit $\Gamma\ll D$ for different choices of
$J/\Gamma$ (colors) and two (solid line) and three (circles, crosses)
impurities. (b) Ratio between the magnitude of the negative and positive
eigenvalues of the low-energy projected interimpurity spin correlators
$\mathbf{S}_{m}\!\cdot\mathbf{S}_{m^{\prime}}$ as a function of $J/\Gamma$
with $\Gamma=D/8$. The solid line is obtained from two-impurity calculations
while the circles and crosses are obtained from three-impurity calculations.}
\end{figure}
\end{minipage}~~~~~%
\begin{minipage}[t]{0.44\columnwidth}%
\begin{table}[H]
\begin{centering}
\begin{tabular}{|c|c|c|}
\hline 
 & $\begin{array}{c}
\vspace{-1em}\\
F_{13}\\
\vspace{-1em}
\end{array}$ & $F_{23}$\tabularnewline
\hline 
\hline 
$\begin{array}{c}
\vspace{-1em}\\
J=\Gamma/2\\
\vspace{-1em}
\end{array}$ & $\frac{1}{\sqrt{2}}\!\!\left(\!\begin{smallmatrix}0.945 & 1.052\\
1.052 & -0.945
\end{smallmatrix}\!\right)$ & $\frac{1}{\sqrt{2}}\!\!\left(\!\begin{smallmatrix}0.60 & \!\!\!+1.28ie^{\!+1.2i}\\
-1.28ie^{\!-1.2i} & \!\!\!-0.60
\end{smallmatrix}\!\right)$\tabularnewline
\hline 
$\begin{array}{c}
\vspace{-1em}\\
J=\Gamma\\
\vspace{-1em}
\end{array}$ & $\frac{1}{\sqrt{2}}\!\!\left(\!\begin{smallmatrix}0.992 & 1.008\\
1.008 & -0.992
\end{smallmatrix}\!\right)$ & $\frac{1}{\sqrt{2}}\!\!\left(\!\begin{smallmatrix}0.83 & \!\!\!+1.15ie^{\!+0.3i}\\
-1.15ie^{\!-0.3i} & \!\!\!-0.83
\end{smallmatrix}\!\right)$\tabularnewline
\hline 
$\begin{array}{c}
\vspace{-1em}\\
J=2\Gamma\\
\vspace{-1em}
\end{array}$ & $\frac{1}{\sqrt{2}}\!\!\left(\!\begin{smallmatrix}1.003 & 0.997\\
0.997 & -1.003
\end{smallmatrix}\!\right)$ & $\frac{1}{\sqrt{2}}\!\!\left(\!\begin{smallmatrix}1.05 & \!\!\!+0.95ie^{\!-0.1i}\\
-0.95ie^{\!+0.1i} & \!\!\!-1.051
\end{smallmatrix}\!\right)$\tabularnewline
\hline 
$\begin{array}{c}
\vspace{-1em}\\
\text{Expected}\\
\vspace{-1em}
\end{array}$ & $\frac{1}{\sqrt{2}}\!\!\left(\begin{smallmatrix}1 & 1\\
1 & -1
\end{smallmatrix}\right)$ & $\frac{1}{\sqrt{2}}\!\!\left(\begin{smallmatrix}1 & +i\\
-i & -1
\end{smallmatrix}\right)$\tabularnewline
\hline 
\end{tabular}\vspace{1.5em}
\par\end{centering}
\caption{Numerically extracted $F$ matrices for different choices of $J/\Gamma$
with $\Gamma=D/8$. The last row displays the expected $F$ matrices
of $\mathrm{SU}\!\left(2\right)_{2}$ anyons. \label{tab:F-matrices}}
\end{table}
\end{minipage}

\end{figure*}

For three impurities, we defined (in the main text) the $2\times2$
matrices $\mathbf{s}_{mm^{\prime}}$ as the projection of the spin
correlators $\mathbf{S}_{m}\!\cdot\mathbf{S}_{m^{\prime}}=\mathbb{1}_{\mathrm{2CK}}\otimes\mathbf{s}_{mm^{\prime}}$
onto the fusion space. We then found that they had one negative (singletlike)
and one positive (tripletlike) eigenvalue, with eigenstates denoted
by $\k{0_{mm^{\prime}}}$ and $\k{1_{mm^{\prime}}}$, respectively.
We now define $\alpha_{mm^{\prime}}$ as (minus) the ratio between
the two eigenvalues
\begin{equation}
\alpha_{mm^{\prime}}\equiv-\frac{s_{mm^{\prime}}^{-}}{s_{mm^{\prime}}^{+}}\quad;\quad\begin{cases}
\mathbf{s}_{mm^{\prime}}\k{0_{mm^{\prime}}}=s_{mm^{\prime}}^{-}\k{0_{mm^{\prime}}}\\
\mathbf{s}_{mm^{\prime}}\k{1_{mm^{\prime}}}=s_{mm^{\prime}}^{+}\k{1_{mm^{\prime}}}
\end{cases}.
\end{equation}
For two impurities we do not have the $\mathbf{s}_{mm^{\prime}}$
matrices, but can still define $\alpha_{12}$ as the ratio between
the positive and negative eigenvalues of $\mathbf{S}_{1}\!\cdot\mathbf{S}_{2}$.
In Fig.\,\ref{fig:EV-ratio} we plot $\alpha_{mm^{\prime}}$ as a
function of the ratio $J/\Gamma$, for both two (blue) and three (red)
impurities and two channels. We observe that for $J\gtrsim\Gamma$
the ratio is minimal. If the two impurities are uncorrelated, i.e.,
$\left\langle \mathbf{S}_{m}\!\cdot\mathbf{S}_{m^{\prime}}\right\rangle _{T\to0}=0$,
then for two channels the two eigenvalues of the low-energy projected
$\mathbf{S}_{m}\!\cdot\mathbf{S}_{m^{\prime}}$ should indeed be equal
(and opposite in sign), i.e., $\alpha_{mm^{\prime}}=1$, as obtained
analytically in Ref.\,\citep{gabayMultiimpurityChiralKondo2022}.
However, even in the chiral case we have trivial correlations in the
bath, which in the strong-coupling regime are then mirrored by interimpurity
correlations. As these correlations should decay as the bath spin-density
correlations, i.e., quadratically in the interimpurity distance, in
Ref.\,\citep{gabayMultiimpurityChiralKondo2022}, which assumes large
separation, they were neglected with respect to the fusion-dependent
contribution, which only decays as the first power of the distance.
However, in the limit of short distances, as in our case, they survive
and $\alpha_{mm^{\prime}}>1$. Still, we understand the regions of
minimal $\alpha_{mm^{\prime}}$ as the best converged regions.

We also define a generalized $F$ matrix
\begin{equation}
F_{mm^{\prime\prime}}=\begin{pmatrix}\bk{0_{mm^{\prime}}|0_{m^{\prime}m^{\prime\prime}}} & \bk{0_{mm^{\prime}}|1_{m^{\prime}m^{\prime\prime}}}\\
\bk{1_{mm^{\prime}}|0_{m^{\prime}m^{\prime\prime}}} & \bk{1_{mm^{\prime}}|1_{m^{\prime}m^{\prime\prime}}}
\end{pmatrix}\quad;\quad m\neq m^{\prime}\neq m^{\prime\prime},
\end{equation}
which describes the basis transformation from the definite fusion
outcomes eigenstates of anyons $m,m^{\prime}$ to those of anyons
$m^{\prime},m^{\prime\prime}$. Thus, $F_{13}$ corresponds to the
standard $F$ matrix defined in Eq.\,(\ref{eq:F-matrix}) of the
main text. $F_{23}$ can be obtained by braiding anyons 1 and 2, transforming
according to $F_{13}$, and then unbraiding. $F_{12}$ can be obtained
by a similar procedure, but it is actually already fully specified
by $F_{12}=F_{23}^{\dagger}F_{13}$.  In Table \ref{tab:F-matrices}
we present the extracted $F$ matrices for different choices of $J/\Gamma$,
as well as the expected value for $\mathrm{SU}\!\left(2\right)_{2}$
anyons. We observe that also in this case we converge for $J\gtrsim\Gamma$,
with the nearest-neighbor related term $F_{13}$ converging at a faster
rate (already for $J=\Gamma$) than $F_{23}$.

\subsection{Spectrum}

In Sec.\,\ref{sec:Finite-Size-Spectrum} we analyze the low-energy
fixed point spectrum, which we can compare with the finite-size spectrum
obtained from the CFT ansatz. The two should agree for $\Lambda\to1$
which takes NRG back to the continuum limit, while larger $\Lambda$
induces ``corrections'' to the NRG spectrum. For $\Lambda=3$ these
corrections become significant, i.e., lead to artificial splitting
which is larger than the level spacing, already at low energy levels,
so that the comparison with the CFT spectrum becomes difficult. Note
that these deviations to not affect ``global'' spectrum properties
such as the impurity entropy. Thus, only for the spectrum presented
in Fig.\,\ref{fig:Spectrum}, we we take smaller $\Lambda=2$. However,
this renders the regime between $D$ and $J,\Gamma$ inaccessible
within our computational resources. In order to overcome this challenge,
we employ a useful NRG trick of taking the physical parameters $J$
and $\Gamma$ to be larger than the sharp cutoff $D$. This guarantees
that already at the first NRG iteration we can discard the high-energy
states of the effective-impurity, as the Wilson-chain couplings, which
are smaller than $D$, cannot mix them with the low-energy states.
We thus completely skip the difficult regime, and allow the RG flow
to correct errors introduced by this artificial choice of parameters.
Tuning the chiral hopping amplitudes $t_{mm}^{\prime}$ in this case,
we find that they deviate significantly from $\Gamma$ and are as
small as $0.4\Gamma$, as shown in Fig.\,\ref{fig:chiral-hopping}.
This is understandable, as their expected value of $\Gamma$ was only
derived in the limit of $D\to\infty$, which is clearly not the case
here.

\subsection{Fine-Tuning to Chirality\label{subsec:Fine-Tuning-to-Chirality}}

\begin{wrapfigure}{o}{0.5\columnwidth}%
\begin{centering}
\vspace{-2em}
\includegraphics[width=0.5\columnwidth]{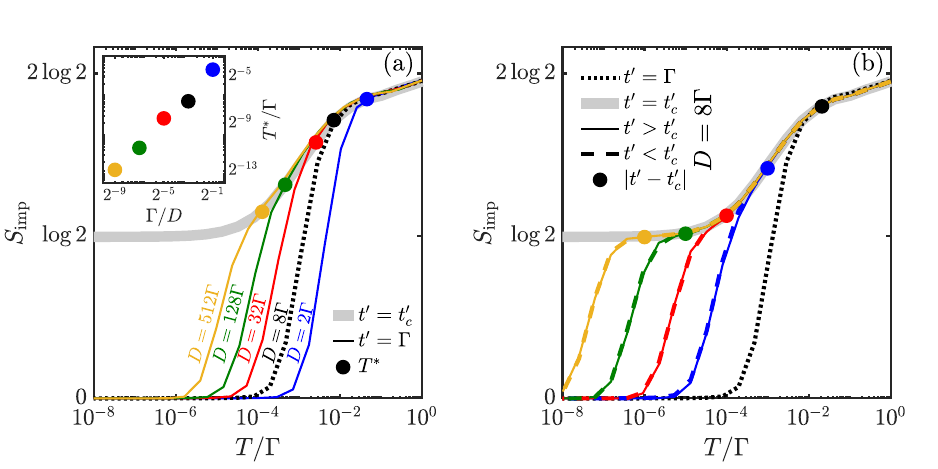}
\par\end{centering}
\begin{centering}
\phantomsubfloat{\label{fig:Impurity-entropy-cutoff}}\phantomsubfloat{\label{fig:Impurity-entropy-fine-tuning}}\vspace{-2.5em}
\par\end{centering}
\caption{Impurity entropy as a function of temperature for two impurities and
two channels, when deviating from the chiral critical point (gray).
All quantities are taken in units of $\Gamma$, with the Kondo coupling
$J=\Gamma/2$, and $\Lambda=6$. (a) Deviations from chirality when
fixing $t^{\prime}=\Gamma$ and varying the sharp cutoff $D$. Large
$D/\Gamma$ requires $N_{K}\protect\leq8000$ kept multiplets in order
to negate truncation effects which shift $t_{c}^{\prime}$. The low-energy
chirality breaking scale $T^{*}$ (circles) is shown to be linear
in $\Gamma/D$ (inset). (b) Deviations from chirality for $D=8\Gamma$
with positive (solid) and negative (dashed) deviations of $t^{\prime}-t_{c}^{\prime}$.
Here $N_{K}\protect\leq2500$ suffices.}
\end{wrapfigure}%

We now return to study the chirality-breaking effects of the sharp
numerical cutoff $D$, and their remedy by numerically fine-tuning
the imaginary hopping amplitudes $t_{mm^{\prime}}^{\prime}$. We focus
on the two-impurity case, so that we have only a single hopping amplitude
$t^{\prime}\equiv t_{12}^{\prime}$, and on two channels, so that
we can significantly crank up the different numerical parameters within
reasonable resources, but the following discussion is equally valid
for more impurities and channels. In the absence of the sharp cutoff,
$t^{\prime}$ is equal to the value of the soft cutoff $\Gamma$,
rendering the system chiral. Fixing $t^{\prime}=\Gamma$ and $J$
(and thus $T_{K}$) with respect to $\Gamma$, in Fig.\,\ref{fig:Impurity-entropy-cutoff}
we plot the impurity entropy for different ratios $D/\Gamma>1$ (approaching
$D\to\infty$). Observe that it follows the universal curve of two
infinitely separated impurities down to some low-energy scale $T^{*}$
which violates chirality. Defining $T^{*}$ by a deviation of 1\%
from the universal curve (circles), we find that it decays linearly
in the ratio between the soft and sharp cutoffs $\Gamma/D$ (inset).
Thus, in the presence of a sharp cutoff and when fixing $t^{\prime}=\Gamma$,
the effective system is chiral only in an intermediate regime between
the two energy scales $T^{*}$ and $D$.

This behavior mimics the effect of the bulk bands (Landau levels),
which we have mostly ignored in this work, choosing to focus on the
chiral edge mode residing in the band gap. The bulk band gap sets
a finite bandwidth for the chiral mode, akin to $D$, while virtual
process through the bulk bands mediate effective RKKY interactions
at a low-energy scale $K$, which violates chirality, akin to $T^{*}$.
However, whereas $T^{*}$ decays linearly with the artificial ratio
between the two cutoffs $\Gamma/D$, the physical $K$ decays exponentially
with both the bulk gap and the interimpurity distance \citep{bloembergenNuclearSpinExchange1955,kurilovichIndirectExchangeInteraction2016,kurilovichIndirectExchangeInteraction2017}.
This difference is not surprising, as in the effective model we have
first taken the limit of an infinite band gap (so that $K\to0$),
then a small interimpurity separation (but still larger than the gap-induced
short-distance scale), and only then reintroduced the artificial cutoff
$D$ out of numerical necessity. Due to the exponential decay of the
physical low-energy scale $K$, it is reasonable to assume that in
an experiment it is negligible, and to thus also eliminate it (and
its equivalent $T^{*}$) in the numerical analysis by tuning $t^{\prime}$
to be equal to $t_{c}^{\prime}$, as we now further elaborate.

In the absence of the cutoff $D$, the dictated $t_{c}^{\prime}=\Gamma$
sits at the critical point of a quantum phase transition of the effective
model, with $T^{*}=0$, so that the system remains chiral down to
arbitrarily small energy scales. The introduction of the sharp cutoff
might shift the critical value $t_{c}^{\prime}$ away from $\Gamma$,
but is not expected to affect the chiral nature of quantum critical
point, as indeed observed numerically. Thus, for a given set of numerical
parameters, and foremost $D$, we can numerically search for $t_{c}^{\prime}$
(there are many valid observables for the search, e.g., $\left\langle \mathbf{S}_{1}\!\cdot\mathbf{S}_{2}\right\rangle _{T\to0}$,
$S_{\mathrm{imp}}^{T\to0}$, and in practice we look at the ground-state
degeneracy of the effective NRG Hamiltonian at some late iteration).
Note that this procedure accounts also for other chirality breaking
effects, such as the NRG discretization and truncation, which also
shift $t_{c}^{\prime}$, although to a lesser extent than the sharp
cutoff (this becomes apparent when the cutoff is very large, so that
its effect is small). Having found $t_{c}^{\prime}$, we can study
deviations from it, as shown in Fig.\,\ref{fig:Impurity-entropy-fine-tuning},
both for $t^{\prime}>t_{c}^{\prime}$ (solid) and $t^{\prime}<t_{c}^{\prime}$
(dashed). Observe that the energy scale of the deviation from the
universal curve ($t^{\prime}=t_{c}^{\prime}$ or two infinitely separated
impurities) is simply $\left|t^{\prime}-t_{c}^{\prime}\right|$ (full
circles).

We thus understand that if we fix $t^{\prime}=\Gamma$, as in Fig.\,\ref{fig:Impurity-entropy-cutoff},
then the low-energy chirality breaking scale is $T^{*}=\Gamma-t_{c}^{\prime}$,
and is consistent with the linear scaling of $1-t_{c}^{\prime}/\Gamma$
in Fig.\,\ref{fig:chiral-hopping}. But we also understand that setting
$t^{\prime}=t_{c}^{\prime}$ eliminates the low-energy chirality breaking
scale, even in the presence of a finite bandwidth. As the physical
low-energy chirality breaking term decays exponentially, we find this
choice better represents an experimental setup.